# The ion-deposited energy formulas and their applications for the interaction of various energetic ions with carbon nanostructures

Li-Ping Zheng[1], Jian-Qing Cao[1], Peng-Yu Ren[1], Guo-Liang Ma[2,1]

[1]Shanghai Institute of Applied Physics, Chinese Academy of Sciences, P.O. Box 800-204, Shanghai 201800, China

[2]Key Laboratory of Nuclear Physics and Ion-beam Application (MOE), Institute of Modern Physics, Fudan University, Shanghai 200433, China

**Abstract**

Based on Monte Carlo (MC) simulations of the interaction of energetic noble gas ions with carbon nanostructures, we have derived the ion-deposited energy formulas and their applications for the above interaction. We also have studied the nuclear point mass effects i.e. the well-known mass ratio effects, and the nuclear point charge effects on the ion-deposited energy then the damage regularities. In particular, we, for first time, find a straight line of the penetrating point threshold energies for energetic noble gas ions. Near this straight line, the nuclear point charge effects dominate the ion-deposited energy then the damage regularities.

## 1. Introduction

At low temperature, low dose and normal incidence, to understand the basic physics of the radiation damage, the stopping and the penetrating for the interaction of energetic noble- gas ions with carbon nanostructures, it is necessary to consider ion irradiations of both the SWCNT (single-walled carbon nanotube) with a shell and the MWCNT (multi-walled carbon nanotube) with multi- shells, as shown in Fig.2 of [1], Fig. 1 of [2] and Fig. 2 of [3] etc. Especially, one (Fig.2 of [1]), two and three shells (Fig.2 of [3] and Fig.4 of [4]) (thin irradiated targets) are suitable for study of the damage mechanisms [1-15].

It is clearly known that the ion- deposited energy plays an important role in studying the damage mechanisms. Therefore, for better understanding of these mechanisms [1-15], it might be informative that we calculate the deposited energies of various- energetic ions in the SWCNT or the MWCNT. At low temperature, low dose and normal incidence, the MD (molecular dynamics) simulation programs with the ZBL (Ziegler- Biersack- Littmark) potentials have been used to study the stopping for the interaction of energetic noble- gas ions with carbon nanostructures [2-3] except Ref. [1] . At low temperature, low dose limit and normal incidence, a static MC (Monte Carlo) simulation program with the Moliere potentials has been used to study the penetrating for the interaction of energetic noble- gas ions with carbon nanostructures [4]. In our opinion, the meaning of the penetrating is very different from that of the stopping. The penetrating means that incident ions penetrate through the same shells [4], but the stopping means that both incident ions and carbon- atom- recoils stop in the same shells [2-3]. Therefore, the penetrating is only the physical behavior of incident ions but the stopping is that of both incident ions and carbon- atom- recoils.

In the present works, we have derived the ion- deposited energy formulas and their applications for the above interaction. As mentioned in sections 1) Ion-nuclear point charge effects and ion- nuclear point mass effects, 2) Dominant ion- nuclear point mass effects, 3) The ion- penetrating point threshold energy, 4) Physical cause of the ion- penetrating point threshold energy line, 5) Dominant ion- nuclear point charge effects, 6) Coordination defect numbers. As analyzed in sections 1), the cross-section $\sigma$ effects ($\sigma = \pi p^2$, Fig. 3 of [1] and Fig. 7 of [3]) belong to the nuclear point charge $Ze$ ones, and 7) in fact, Ref. [1] is a paper of studying the penetrating ($E_0 > 0.3 keV$), the penetrating threshold energy ($E_0 = 0.3 keV$) then the coordination defect numbers ($E_0 > 0.3 keV$).

**2. Ion- deposited energy formulas**

If the incident energy $E_0$ is high enough, the incident ion easily penetrates through the SWCNT or the MWCNT while it becomes the penetrating ion with the $E_{penetrating}$ kinetic energy, so $E_0 - E_{penetrating}$ equals the ion- deposited energy in the SWCNT or the MWCNT. We have calculated the ion- deposited energy $E_0 - E_{penetrating}$ to distinguish between penetrating and stopped ions [4]. Obviously, if kinetic energy $E_{penetrating} > 0$, i.e. $E_0 > E_0 - E_{penetrating}$ formula (1), the incident ion becomes the penetrating one; if kinetic energy $E_{penetrating} = 0$, i.e. $E_0 = E_0 - E_{penetrating}$ formula (2), the incident ion becomes the stopped one.

Under various energetic ion irradiations, at the same incident energy, i.e. this $E_0$ kinetic energy keeps constant, in a binary collision model based on the Moliere potential for the ion-atom (or atom-atom) interaction, we derive an ion- deposited energy formula (formula 7 of Ref. [4]) for the irradiated MWCNT (or SWCNT) here, that is

$$E_0 - E_{penetrating} = (4Mm/(M+m)^2) \sum_{i=0}^{n-1} E_i \sin^2(\theta_{i+1}/2) \qquad (3)$$

Where, the final ion- kinetic- energy $E_{penetrating}$ in $n$ collisions; the $E_0 - E_{penetrating}$ ion- deposited energy equals the product of the $4Mm/(M+m)^2$ ratio and the $\sum_{i=0}^{n-1} E_i \sin^2(\theta_{i+1}/2)$ angle- correlated energy. Evidently, this formula studies relation between the nuclear point charge effects and its mass ones on the $E_0 - E_{penetrating}$ ion- deposited energy.

As mentioned above, if kinetic energy $E_{penetrating} = 0$, i.e. $E_0 = E_0 - E_{penetrating}$ formula (2), the incident ion becomes the stopped one in the MWCNT (or SWCNT). We combine formula (2) and formula (3), that is

$$E_0 = E_0 - E_{penetrating} = (4Mm/(M+m)^2) \sum_{i=0}^{n-1} E_i \sin^2(\theta_{i+1}/2) \qquad (4)$$

## 3. Application

**1) Ion-nuclear point charge effects and ion- nuclear point mass effects**

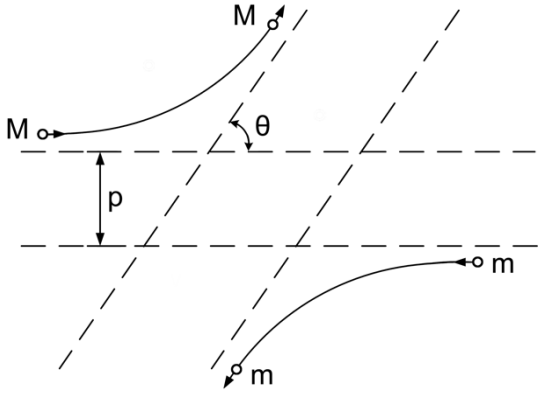

**Fig. 1** Trajectories of the energetic ion and a carbon atom in the interaction potentials of nuclear point charges, in the center of mass system. Here, $M$ and $m$ are nuclear point masses of the ion and a carbon atom, $p$ is the impact parameter and $\theta$ is the ion- scattered angle.

As shown in Fig. 1 and as analyzed in Ref. [4], energetic ions and carbon atoms can be seen as particles with nuclear point charge and nuclear point mass, in particle collisions. The well- known $4Mm/(M+m)^2$ ratio is based on the conservation of energy and that of momentum. However, the $p$ impact parameter and then the $\theta$ ion- scattered angle necessarily exist in the interaction potentials of nuclear point charges. Therefore, the $4Mm/(M+m)^2$ effects belong to nuclear point mass ones, i.e. potential- independent effects, but the $p$ impact parameter effects and the $\theta$ ion- scattered angle ones belong to nuclear point charge $Ze$ effects ($Z$ is the proton number of the ion, and $e$ is the charge on a proton.), i.e. potential- dependent ones. Here, $M$ and $m$ are representative of the energetic ion mass and the carbon atom one, respectively. Evidently, the cross- section $\sigma$ effects ($\sigma = \pi p^2$, Fig. 3 of [1] and

Fig. 7 of [3]) belong to the nuclear point charge $Ze$ ones. Of course, the angle- correlated energy $\sum_{i=0}^{n-1} E_i \sin^2(\theta_{i+1}/2)$ effects belong to the nuclear point charge $Ze$ ones.

**2) Dominant ion- nuclear point mass effects**

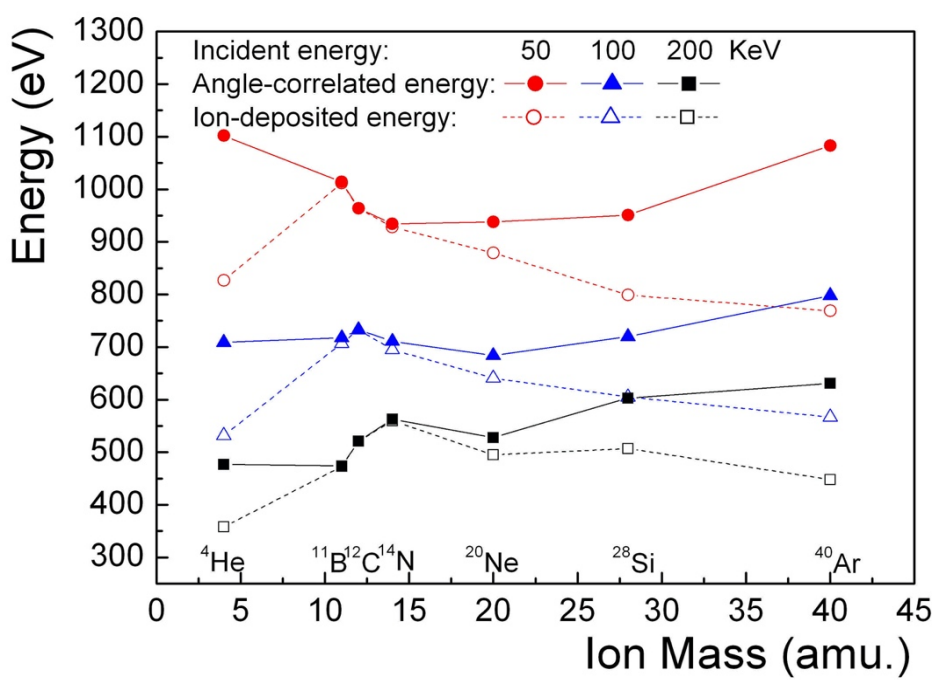

**Fig. 2** The ion- deposited energy $E_0 - E_{penetrating}$ (open) and the angle- correlated energy $\sum_{i=0}^{n-1} E_i \sin^2(\theta_{i+1}/2)$ (solid) as functions of the ion mass for $^4$He, $^{11}$B, $^{12}$C, $^{14}$N, $^{20}$Ne, $^{28}$Si and $^{40}$Ar ion- irradiations of a MWCNT composed of (10,10), (15,15) and (20,20) SWCNTs, at $50$ (circles), $100$ (triangles) and $200 keV$ (squares) incident energies.

Fig. 2 shows that under hundreds $keV$ $^4$He, $^{11}$B, $^{12}$C, $^{14}$N, $^{20}$Ne, $^{28}$Si and $^{40}$Ar ion- irradiations, the ion- deposited energy $E_0 - E_{penetrating}$ maximizes, while the ion mass has intermediate mass values, such as $^{11}$B, $^{12}$C and $^{14}$N ion masses. Namely, the

ion- deposited energy $E_0 - E_{penetrating}$ maximizes for $50keV$ $^{11}$B, $100keV$ $^{12}$C and $200keV$ $^{14}$N ion- irradiations of 3 shells. In our opinion, for the MWCNT irradiated by hundreds $keV$ ions, if the $4Mm/(M+m)^2$ effects, i.e. the nuclear point mass effects, dominate over the nuclear point charge effects i.e. the angle- correlated energy $\sum_{i=0}^{n-1} E_i \sin^2(\theta_{i+1}/2)$ effects, in Fig. 2 the ion- deposited energy $E_0 - E_{penetrating}$ has a maximum, while the ion mass equals an intermediate mass value, such as $^{11}$B, $^{12}$C or $^{14}$N ion masses etc. Corresponding to Fig. 2, Fig. 4 of [4] has shown that the $4Mm/(M+m)^2$ effects i.e. the nuclear point mass effects, dominate the damage regularities.

**3) The ion- penetrating point threshold energy.**

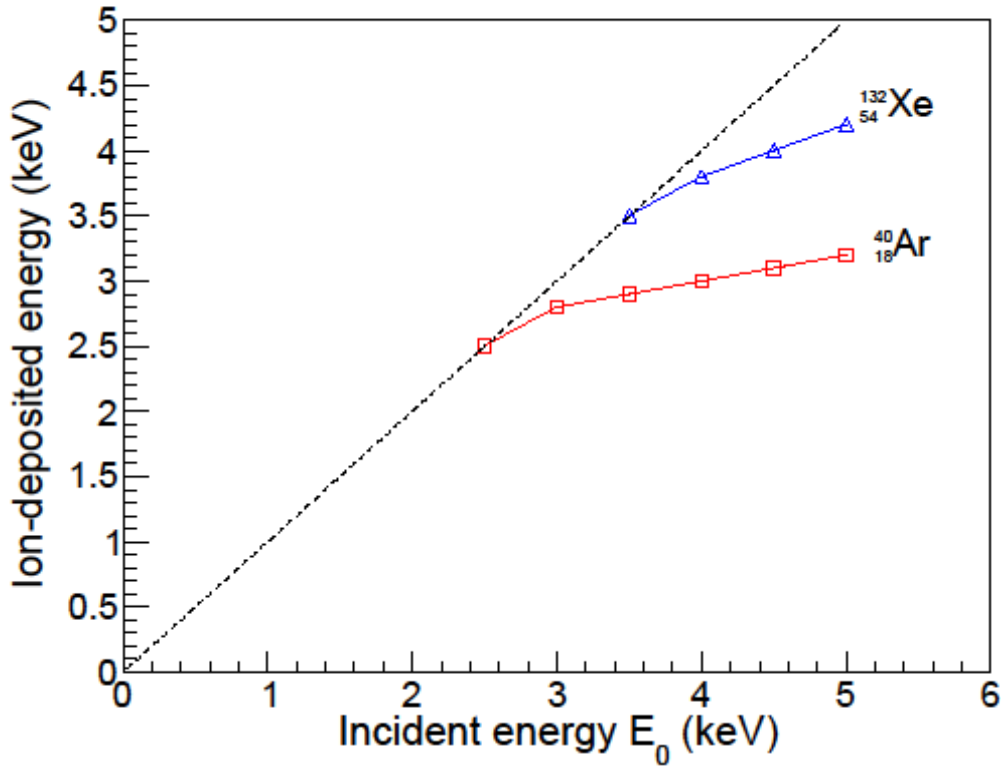

**Fig. 3** The ion- deposited energy $E_0 - E_{penetrating}$ in the zigzag MWCNT with 10

shells as a function of the incident ion energy $E_0$ for various incident ions, such as $^{40}Ar^{18}$ or $^{132}Xe^{54}$ ions etc.

In Fig. 3, we show a straight- line of the penetrating point threshold energies for various incident ions, and two simple penetrating energy curves for $^{40}Ar^{18}$ and $^{132}Xe^{54}$ ion irradiations of 10 shells, in two dimensions, i.e. the incident energy $E_0$ and the ion- deposited energy dimensions. Two simple curves cross a straight- line at incident energies being 2.5 $keV$ point for $^{40}Ar^{18}$ ions and 3.5 $keV$ point for $^{132}Xe^{54}$ ions, i.e. $^{40}Ar^{18}$ penetrating point threshold energy equals 2.5 $keV$ and $^{132}Xe^{54}$ penetrating point threshold energy equals 3.5 $keV$.

Indeed, as indicated in kinetic energy of the penetrating ion $E_{penetrating} = 0$ i.e. $E_0 = E_0 - E_{penetrating}$ formula (2), Fig .3 shows the straight energy line of various incident ions stopping, such as $^{40}Ar^{18}$ or $^{132}Xe^{54}$ incident ions stopping etc. As indicated in kinetic energy of the penetrating ion $E_{penetrating} > 0$ i.e. $E_0 > E_0 - E_{penetrating}$ formula (1), Fig .3 also shows that two simple penetrating energy curves slowly decrease then cross this straight- line with help of formula (2). Namely, combination of formulas (1) and (2) can determine the ion- penetrating point threshold energy.

In order to systematically research into the penetrating- dependence on a shell, multi- shells and various incident ions, we give the calculation- results simply in Fig. 3, in detail in Fig.4, and briefly in Table 1. For the same shell, the penetrating point threshold energy increases from light to heavy noble gas ions. For the same noble gas ion, the penetrating point threshold energy of three shells is three times of the one of a shell, but the penetrating point threshold energy of two shells is two times of the one of a shell (Fig. 3, Fig. 4 and Table 1).

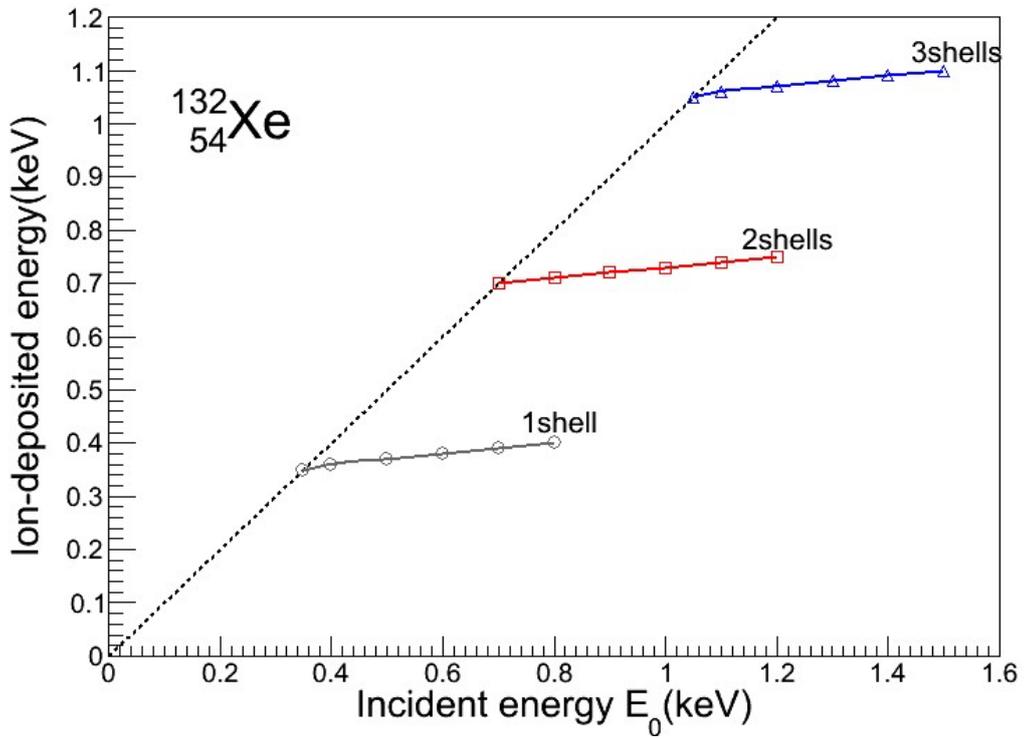

**Fig. 4a** The ion- penetrating point threshold energy as a function of the shell- number under energetic $^{132}Xe^{54}$ ion irradiations of a shell or multi- shells.

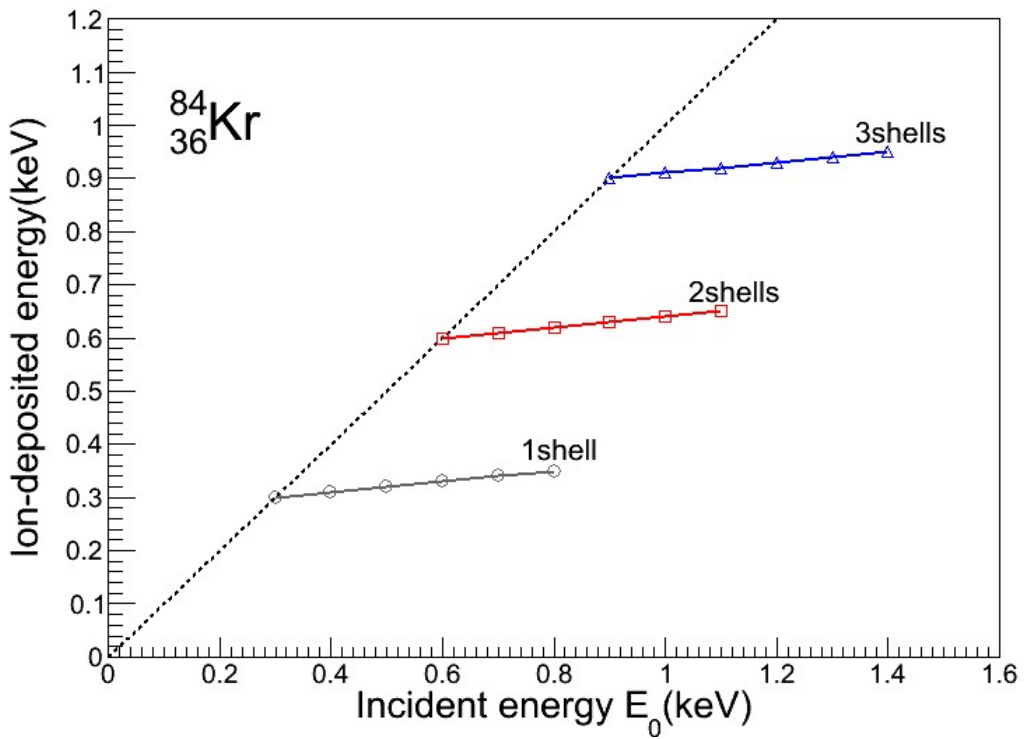

**Fig. 4b** The ion- penetrating point threshold energy as a function of the shell- number under energetic $^{84}Kr^{36}$ ion irradiations of a shell or multi- shells.

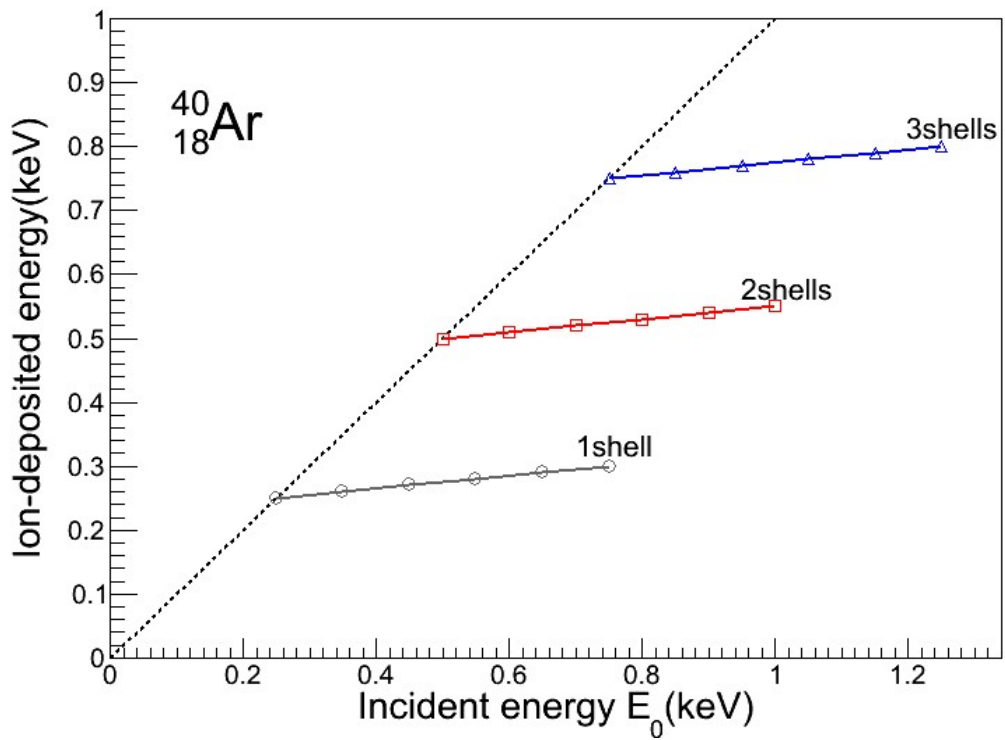

**Fig. 4c** The ion- penetrating point threshold energy as a function of the shell- number under energetic $^{40}Ar^{18}$ ion irradiations of a shell or multi- shells.

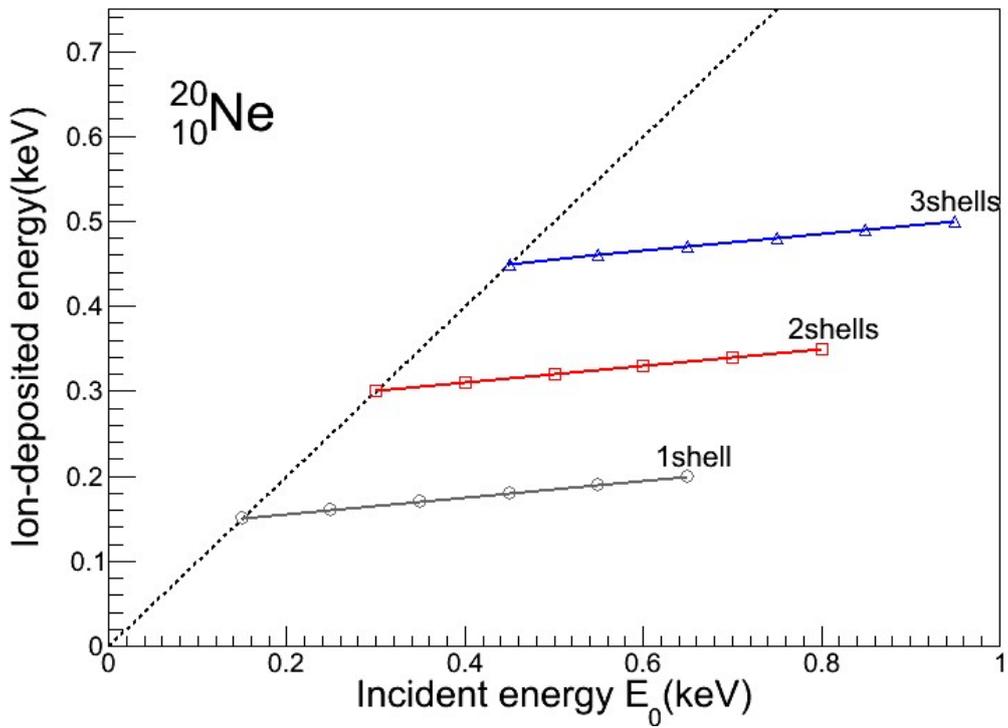

**Fig. 4d** The ion- penetrating point threshold energy as a function of the shell- number under energetic $^{20}Ne^{10}$ ion irradiations of a shell or multi- shells.

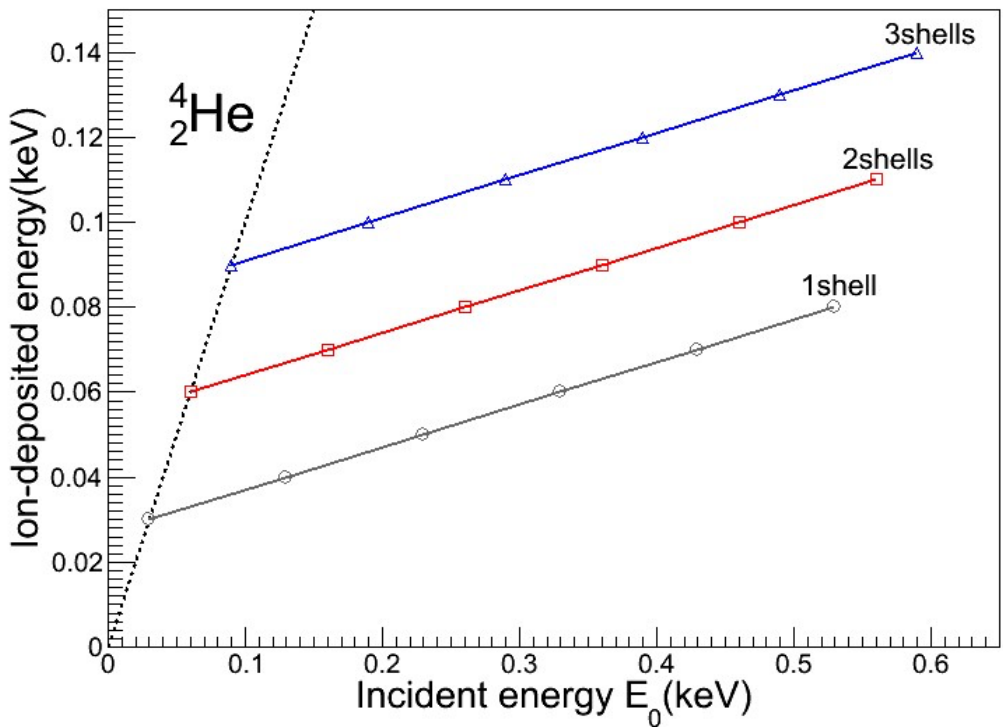

**Fig. 4e** The ion- penetrating point threshold energy as a function of the shell- number

under energetic $^4$He$^2$ ion irradiations of a shell or multi- shells.

**Fig. 4(a, b, c, d and e)** The ion- penetrating point threshold energy as a function of the shell- number under energetic $^4$He$^2$, $^{20}$Ne$^{10}$, $^{40}$Ar$^{18}$, $^{84}$Kr$^{36}$, $^{132}$Xe$^{54}$ ion irradiations of a shell or multi- shells.

**Table 1** The ion- penetrating point threshold energy as a function of the shell- number under energetic noble gas ion irradiations of a shell or multi- shells.

| | Ion- penetrating point threshold energy （$keV$） | | |
|---|---|---|---|
| | 1 shell | 2 shells | 3 shells |
| $^4$He$^2$ | 0.03 | 0.06 | 0.09 |
| $^{20}$Ne$^{10}$ | 0.15 | 0.30 | 0.45 |
| $^{40}$Ar$^{18}$ | 0.25 | 0.50 | 0.75 |
| $^{84}$Kr$^{36}$ | 0.30 | 0.60 | 0.90 |
| $^{132}$Xe$^{54}$ | 0.35 | 0.70 | 1.05 |

**4) Physical cause of the ion- penetrating point threshold energy straight- line**

In Fig. 3 or 4, we show a straight- line of the penetrating point threshold energies for various incident ions. Note that if Fig. 3 or 4 shows a square, the straight line is the one of equally- divided- angle. In formula (4), formula $E_0 = (4Mm/(M+m)^2)\sum_{i=0}^{n-1} E_i \sin^2(\theta_{i+1}/2)$ indicates that the $4Mm/(M+m)^2$ effects and the $\sum_{i=0}^{n-1} E_i \sin^2(\theta_{i+1}/2)$ angle- correlated energy effects cancel each other, while the incident energy $E_0$ is conservative, i.e. it keeps constant. Correspondently, the formula $E_0 = E_0 - E_{penetrating}$ indicates that the incident ion becomes the stopped one. In short, because two nuclear point effects cancel each other, there is a straight- line of the penetrating point threshold energies for various incident ions.

**5) Dominant ion- nuclear point charge effects**

Fig. 4 systematically shows the ion- deposited energy as a function of the shell-number under energetic $^4$He$^2$, $^{20}$Ne$^{10}$, $^{40}$Ar$^{18}$, $^{84}$Kr$^{36}$, $^{132}$Xe$^{54}$ ion irradiations of a shell or multi-shell, near the ion-penetrating point threshold energy line. For the same shell, as $Ze$ increases, the ion-deposited energy increases, i.e. the ion-nuclear point charge effects ($Ze$ effects) dominate the ion-deposited energy. Otherwise, for the same noble gas ion, the ion-deposited energy increases from one shell to three shells.

**6) Coordination defect numbers**

For the given collision system, such as $^4$He$^2$, $^{20}$Ne$^{10}$, $^{40}$Ar$^{18}$, $^{84}$Kr$^{36}$  $^{132}$Xe$^{54}$ irradiations of the zigzag SWCNT or MWCNT etc, a run consists of an incident ion normally and randomly entering a new 'undamaged' zigzag SWCNT or MWCNT, so low-limit of incident doses (or fluences), i.e. the zero-dose (or the zero-fluence), should be realized [12，13]. In other words, MC simulations of the coordination defect numbers of irradiated SWCNT or MWCNT have been performed under the zero-dose.

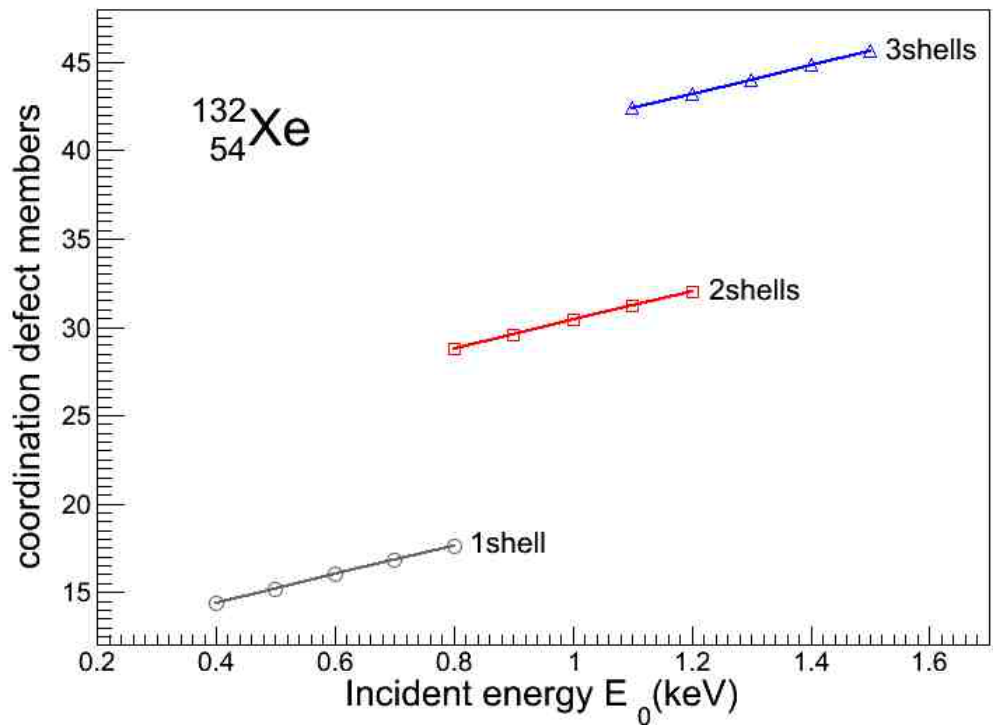

**Fig. 5a** The coordination defect numbers a function of the shell- number under energetic $^{132}Xe^{54}$ ion irradiations of a shell or multi- shells，near the ion- penetrating point threshold energy straight- line.

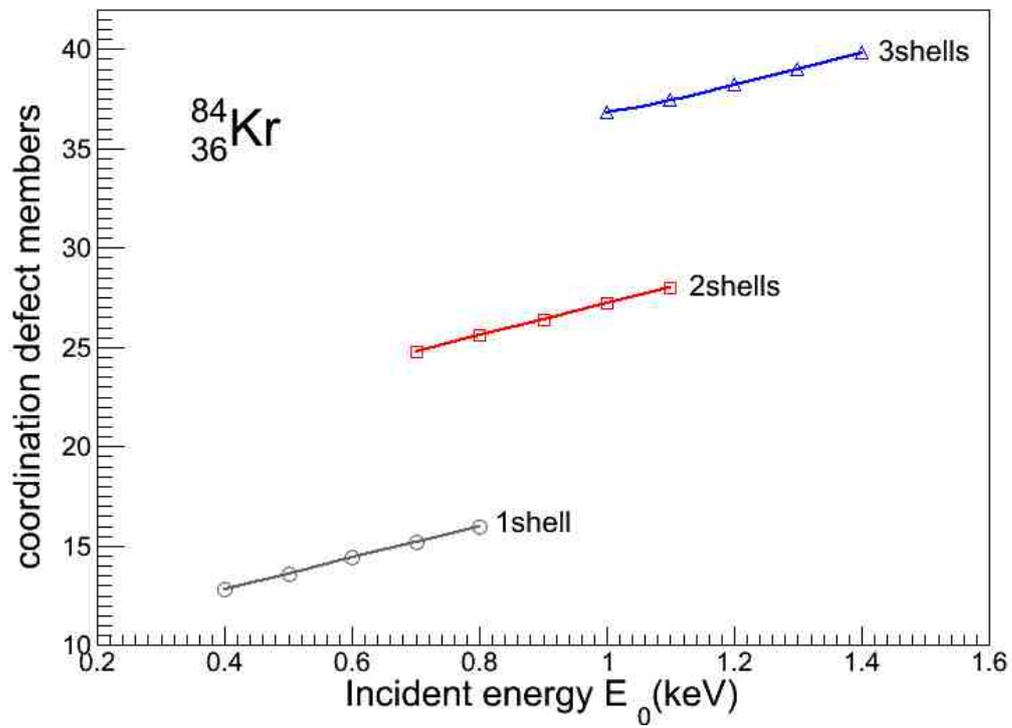

**Fig. 5b** The coordination defect numbers a function of the shell- number under energetic $^{84}Kr^{36}$ ion irradiations of a shell or multi- shells，near the ion- penetrating point threshold energy straight- line.

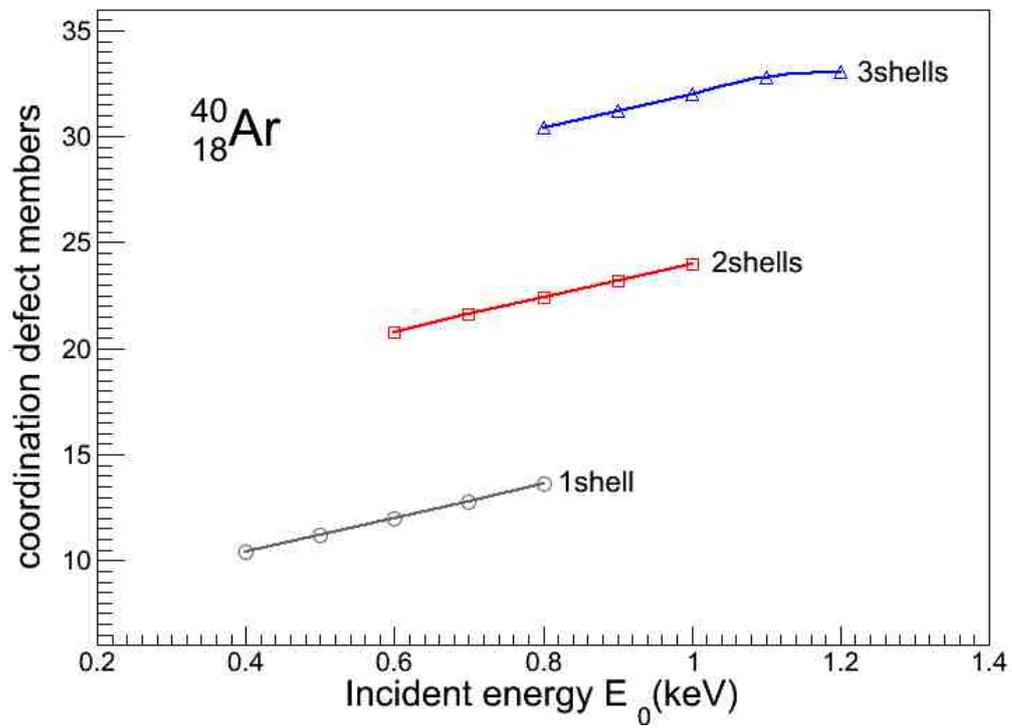

**Fig. 5c** The coordination defect numbers a function of the shell- number under energetic $^{40}Ar^{18}$ ion irradiations of a shell or multi- shells，near the ion- penetrating point threshold energy straight- line.

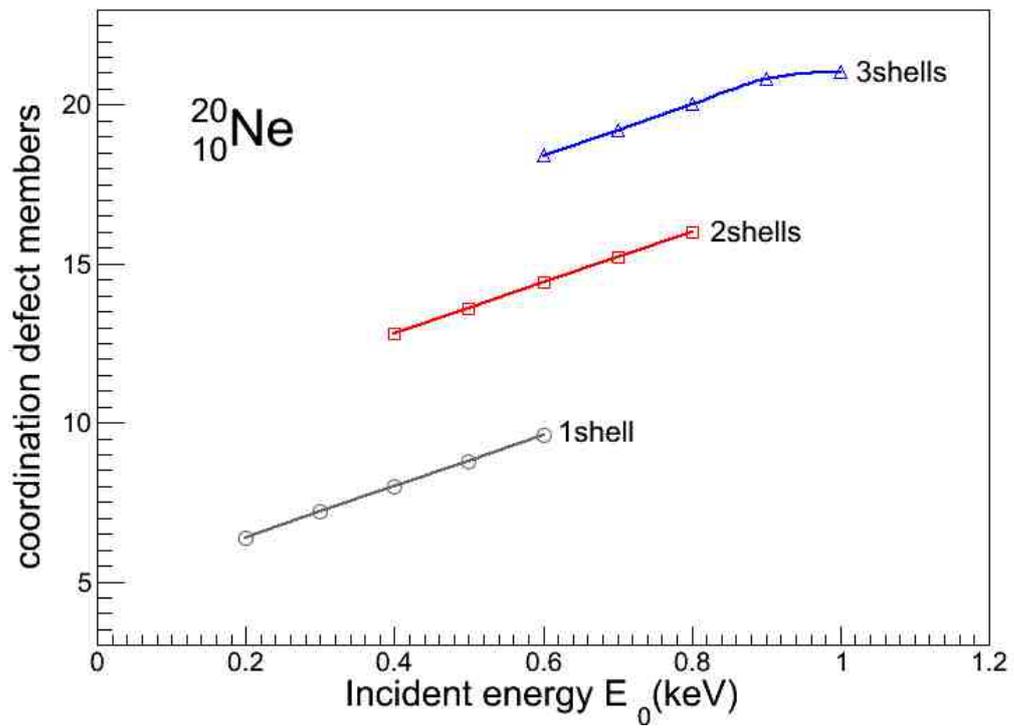

**Fig. 5d** The coordination defect numbers a function of the shell- number under energetic $^{20}Ne^{10}$ ion irradiations of a shell or multi- shells, near the ion- penetrating point threshold energy straight- line.

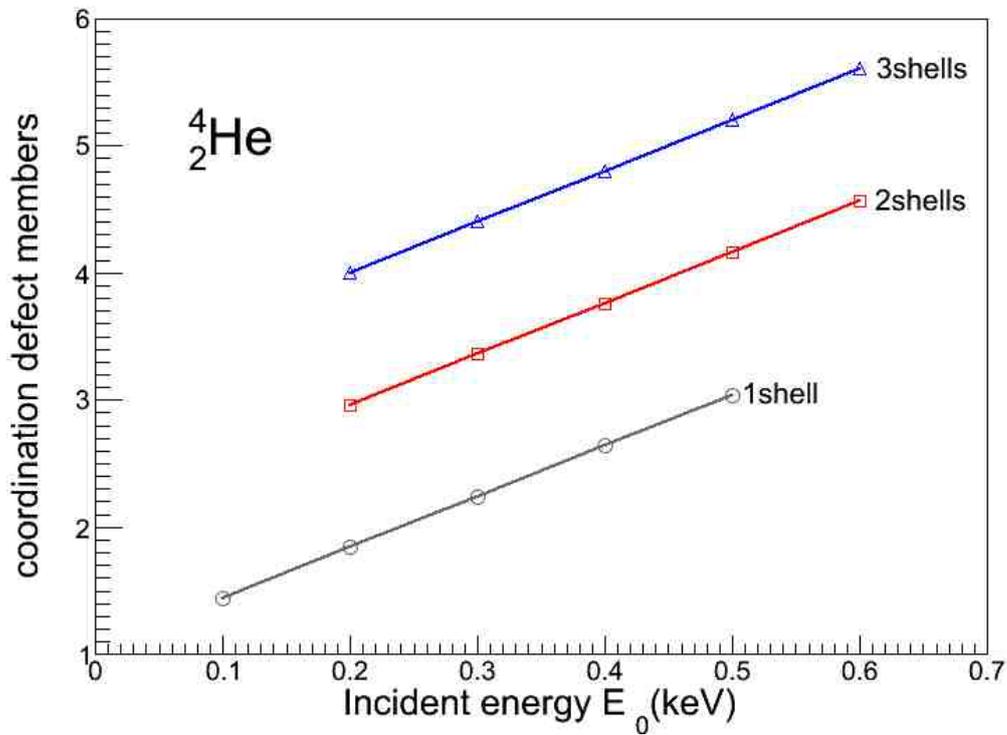

**Fig. 5e** The coordination defect numbers a function of the shell- number under energetic $^4He^2$ ion irradiations of a shell or multi- shells，near the ion- penetrating point threshold energy straight- line.

**Fig. 5(a, b, c, d, and e)** The coordination defect numbers a function of the shell- number under energetic $^4He^2$, $^{20}Ne^{10}$, $^{40}Ar^{18}$, $^{84}Kr^{36}$, $^{132}Xe^{54}$ ion irradiations of a shell or multi- shells，near the ion- penetrating point threshold energy straight- line.

Corresponding to Fig. 4, Fig. 5 systematically shows the coordination defect numbers as a function of the shell- number under energetic $^4He^2$, $^{20}Ne^{10}$, $^{40}Ar^{18}$, $^{84}Kr^{36}$, $^{132}Xe^{54}$ ion irradiations of a shell or multi- shell, near the ion- penetrating point threshold energy straight line. For the same shell, as $Ze$ increases, the coordination defect numbers increase, i.e. the ion- nuclear point charge effects ($Ze$ effects) dominate the coordination defect numbers. Otherwise, for the same noble gas ion, the coordination defect numbers increase from one shell to three shells.

**7) MC and MD simulation result comparisons**

The MD simulation works [1] have found (page 19) that various ions (except Helium ions) lose on the average 0.3 $keV$ of their kinetic energy when penetrating a SWCNT. As shown in Fig. 2 of [1], as $Ze$ increases, the coordination defect numbers increase for $E_0 > 0.3 keV$. In fact, Ref. [1] is a paper of studying the penetrating ( $E_0 > 0.3 keV$ ), the penetrating threshold energy ( $E_0 = 0.3 keV$ ) then the coordination defect numbers ( $E_0 > 0.3 keV$ ) regularities. These MD simulation results [1] including 1 shell of Fig. 2 of [3] are similar to our MC ones of 6).

**4. Summary**

Based on Monte Carlo (MC) simulations of the interaction of energetic noble gas ions with carbon nanostructures, we have derived the ion- deposited energy formulas and their applications for the above interaction. We also have studied the nuclear point mass effects i.e. the well known $4Mm/(M+m)^2$ ones, and the nuclear point charge ones i.e. $Ze$ ones on the ion- deposited energy then the damage regularities. In particular, we, for first time, find a straight- line of the penetrating point threshold energies for energetic noble gas ions. Near this straight- line, the nuclear point charge effects i.e. $Ze$ ones dominate the ion- deposited energy then the damage regularities. Otherwise, for the same noble gas ion, the damage increases from one shell to three shells.

**Acknowledgements**

Supported by National Basic Research Program of China (973 Program) 2010CB832903, National Natural Science Foundation of China No 11175232 and No 11175235.

**References**

[1] J. Pomoell, A.V. Krasheninnikov, K. Nordlund, J. Keincnen, Nucl. Instrum. Methods **2003**, B 206, 18.

[2] A.V. Krasheninnikov, K. Nordlund, J. Keincnen, Appl. Phys. Lett. **81 (2002)** 1101.

[3] J. Pomoell, A.V. Krasheninnikov, K. Nordlund, J. Keincnen, J. Appl. Phys. **96 (2004)** 2864.

[4] Li-Ping Zheng, Long Yan, Zhi-Yong Zhu, Guo-Liang Ma. Appl. Phys. A (**2016**) 122:222.

[5] P. Sigmund, << Particle Penetration and Radiation Effects: General Aspects and Stopping of Point Charge>>, Springer，2006.

[6] P. Sigmund, Phys. Rev. **184 (1969)** 383.

[7] Biersack, J. P.; Haggmark, L. G. Nucl. Instrum. Methods **1980**, 174, 257.

[8] J.F. Ziegler, J.P. Biersack, and U. Littmark, The Stopping and Range of Ions in Matter (Pergamon, New York, 1985)

[9] L.P. Zheng, Z.Y. Zhu, Y. Li, D.Z. Zhu, H.H. Xia, J. Phys. Chem. C **112 (2008)** 15204.

[10] L.P. Zheng, R.S. Li, X.Q. Xia, M.Y. Li, Appl. Phys. A **61 (1995)** 419.

[11] L.P. Zheng , Y.G. Ma, J.G. Han, D.X. Li, X.R. Zhang, Phys. Lett. A **324 (2004)** 211.

[12] L.P. Zheng, Nucl. Instr. and Meth. B **160 (2000)** 29.

[13] L.P. Zheng et al, Appl. Surf. Sci. **143 (1999)** 215.

[14] A.V. Krasheninnikov, K. Nordlund, J. Keincnen, Phys. Rev. B **63 (2001)** 245405.

[15] A.V. Krasheninnikov, K. Nordlund, J. Appl. Phys. **107 (2010)** 071301.